\documentclass{elsart}

\usepackage{epsfig}

\usepackage{amssymb}

\begin{document}

\begin{frontmatter}



\title{Searching for higher-dimensional wormholes with noncommutative geometry}

\author[label1]{Farook Rahaman}\footnote{Corresponding author.}\ead{rahaman@iucaa.ernet.in},
\author[label2]{P.K.F. Kuhfittig}\ead{kuhfitti@msoe.edu},
\author[label3]{Saibal Ray}\ead{saibal@iucaa.ernet.in},
\author[label4]{Safiqul Islam}\ead{sofiqul001@yahoo.co.in}
\address[label1]{Department of Mathematics, Jadavpur University,
Kolkata 700 032, West Bengal, India}
\address[label2]{Department of Mathematics,
Milwaukee School of Engineering, Milwaukee, Wisconsin 53202-3109,
USA}
\address[label3]{Department of Physics, Government College of Engineering and Ceramic
Technology, Kolkata 700 010, West Bengal, India}
\address[label4]{Department of
Mathematics, Jadavpur University, Kolkata 700 032, West Bengal,
India}

\begin{abstract}
Noncommutative geometry, an offshoot of string theory, replaces
point-like structures with smeared objects and has recently been
extended to higher dimensions.  The purpose of this paper is to
obtain wormhole solutions with this extended noncommutative
geometry as a background. It is found through this investigation
that wormhole solutions exist in the usual four, as well as in
five dimensions, but they do not exist in higher-dimensional
spacetimes.
\end{abstract}

\end{frontmatter}

\section{Introduction}
The extension of general relativity to higher dimensions was
motivated in part by studies of the early Universe.  While it is
generally believed that in the present Universe the extra spatial
dimensions have become compactified, their very existence has led
to many investigations in various areas. Rahaman et al.
\cite{RRKS09} investigated whether the usual solar system tests
are compatible with the existence of higher spatial dimensions.
Other studies involved the motion of test particles \cite{LO00},
solar-system effects in a Scharzschild-de Sitter spacetime
\cite{KKL02}, as well as numerous other phenomena
\cite{lI05a,lI05b}.

Many studies assume only one extra spatial dimension when
discussing higher-dimensional cases
\cite{pW83,tF87,BBC90,CB90,RS99a,RS99b,jP03,sR06,pK06}. In this
regard we would like to have a special mention of some of our
recent works \cite{sR06,pK06}. Ray \cite{sR06} has considered
static spherically symmetric charged dust corresponding to $(n +
2)$ dimensional Einstein-Maxwell spacetime and shown that for $n =
3$, the expression for gravitational mass corresponds to the mass
given by Bonnor \cite{bon1960} and Cohen and Cohen \cite{cc1969}
which, in turn, confirms the identification of mass as the fifth
dimension by Ponce de Leon \cite{leo2003}. In the later work,
Kuhfittig \cite{pK06} assumes that the $3$-brane is a de Sitter
space and there is only one extra spatial dimension, assumed to be
time dependent. It is proposed in this work that the cosmological
inflation of the $3$-brane may provide a possible explanation for
the collapse of the extra dimension, as well as for the energy
stored in the resulting curled-up dimension.

An important outcome of string theory (which assumes extra
dimensions) is the notion that coordinates may become
noncommutative operators in a $D$-brane \cite{eW96,SW99}. The
consequence is a fundamental discretization of spacetime due to
the commutator
$[\textbf{x}^{\mu},\textbf{x}^{\nu}]=i\theta^{\mu\nu}$, where
$\theta^{\mu\nu}$ is an antisymmetric matrix, similar to the way
that the Planck constant $\hbar$ discretizes phase space
\cite{aG05}.  This noncommutative geometry is an intrinsic
property of spacetime that does not depend on particular features
such as curvature.  Moreover, it was pointed out in Ref.
\cite{SS03} that noncommutativity replaces point-like structures
by smeared objects, thereby eliminating the divergences that
normally appear in general relativity.  This smearing can be
modeled by the use of the Gaussian distribution of minimal length
$\sqrt{\theta}$ instead of the Dirac delta function.  So the
energy density of the static and spherically symmetric smeared and
particle-like gravitational source has the form \cite{NSS06}
\begin{equation}
  \rho(r)=\frac{M}{(4\pi\theta)^{3/2}}e^{-r^2/4\theta}.
\end{equation}
The mass $M$ could be a diffused centralized object such as a
wormhole \cite{LG04}. The Gaussian source has also been used by
Sushkov \cite{sS05} to model phantom-energy supported wormholes,
as well as by Nicolini and Spalluci \cite{NS10} for the purpose of
modelling the physical effects of short-distance fluctuations of
noncommutative coordinates in the study of black holes. Galactic
rotation curves inspired by a noncommutative-geometry background
are discussed in one of our very recent works \cite{RKCUR12}. The
stability of a particular class of thin-shell wormholes in
noncommutative geometry is analyzed elsewhere \cite{pK12}.

Therefore, the purpose of this paper is to obtain wormhole
solutions within the framework of noncommutative geometry, first
by assuming the usual four dimensions in general relativity and
then by adding one extra spatial dimension.  It was subsequently
discovered, however, that for dimensions above five, no wormhole
solutions exist, pointing out the danger of assuming only one
extra spatial dimension.

\section{Basic Equations}
To describe a spherically symmetric wormhole spacetime  in higher
dimension, we take the metric to be
\begin{equation}
     ds^2=  - e^{\nu(r)} dt^2+ e^{\lambda(r)} dr^2+r^2 d\Omega_n^2,
   \label{Eq3}
   \end{equation}
where the line element ${d{\Omega}_n}^2$ on the unit $n$-sphere
is given by
\begin{eqnarray}
{d{\Omega}_n}^2 = d {{{\theta}_1} ^2} + {sin}^2 {\theta}_1 d
{{{\theta}_2} ^2}+ {sin}^2 {\theta}_1 {sin}^2 {\theta}_2 d
{{{\theta}_3} ^2} +..................+\prod_{i=1}^{n-1} {sin}^2
{\theta}_i d{{\theta}_n} ^2.
\end{eqnarray}

The most general energy momentum tensor compatible with static
spherically symmetry is
\begin{equation}
T_\nu^\mu=  ( \rho + p_r)u^{\mu}u_{\nu} - p_r g^{\mu}_{\nu}+
            (p_t -p_r )\eta^{\mu}\eta_{\nu} \label{eq:emten}
\end{equation}
with $u^{\mu}u_{\mu} = - \eta^{\mu}\eta_{\mu} = 1$.

The Einstein equations are
\begin{equation}\label{E:Einstein1}
e^{-\lambda}
\left(\frac{n \lambda^\prime}{2 r} - \frac{n(n-1)}{2r^2}
\right)+\frac{n(n-1)}{2r^2}= 8\pi \rho,
\end{equation}
\begin{equation}\label{E:Einstein2}
e^{-\lambda}
\left(\frac{n(n-1)}{2r^2}+\frac{n
\nu^\prime}{2r}\right)-\frac{n(n-1)}{2r^2}= 8\pi p_r ,
\end{equation}
\begin{eqnarray}\label{E:Einstein3}
\frac{1}{2} e^{-\lambda} \left[\frac{1}{2}(\nu^\prime)^2+
\nu^{\prime\prime} -\frac{1}{2}\lambda^\prime\nu^\prime +
\frac{(n-1)}{r}({\nu^\prime- \lambda^\prime})
+\frac{(n-1)(n-2)}{r^2}\right] \nonumber \\
-\frac{(n-1)(n-2)}{2r^2}=8\pi p_t.
\end{eqnarray}
In higher dimensions, the energy density of the static and
spherically symmetric smeared and particle-like gravitational
source having a minimal spread Gaussian profile is taken as
\cite{eS08}
\begin{equation}\label{E:density1}
\rho = \frac{M}{(4 \pi \theta)^{{(n+1)/2}}}~{exp}
\left(-\frac{r^2}{4\theta}\right),
\end{equation}
where $M$ is the total mass of the source which  is diffused
throughout a region of linear dimension $\sqrt{\theta}$ due to the
uncertainty.

\section{Solutions}
 We are going to assume a constant redshift function for our
 model, the so-called zero-tidal force solution \cite{MT88},
 to make the wormhole traversable by humanoid travelers.  In
 other words, we have
\begin{equation}\label{E:redshift}
 \nu = \nu_0,
\end{equation}
where $\nu_0$ is a constant.

Using Eq. (\ref{E:redshift}), we can rewrite the field equations
(\ref{E:Einstein1})-(\ref{E:Einstein3}) in terms of the shape
function $b(r)$, where $ b(r) = r(1-e^{-\lambda}) $:

\begin{equation}\label{E:rho}
 \frac{n b^\prime}{2r^2} + \frac{n(n-2)b}{2r^3}
 = \frac{ 8 \pi M}{(4 \pi \theta)^{\frac{n+1}{2}}}~{exp}
\left(-\frac{r^2}{4\theta}\right),
\end{equation}
\begin{equation}\label{E:radial}
8 \pi p_r = - \frac{n(n-1) b}{2r^3} ,
\end{equation}
\begin{equation}\label{E:transverse}
8 \pi p_t =   \frac{(3-n)(n-1) b}{2r^3} - \frac{(n-1)
b^\prime}{2r^2}.
\end{equation}

From Eq. (\ref{E:rho}), we get the following solution for the
shape function:

$b(r) =  \frac{16 \pi}{n r^{n-2}(4 \pi \theta)^{(n+1)/2}
}\frac{\left[ 2^n \left( \frac{1}{\theta}\right)^{-(n+1)/2}\left\{
r^{n+1}2^{(1-n)/2}\left( \frac{1}{\theta}\right)^{(n+1)/2}\left(
\frac{r^2}{\theta}\right)^{-(n+1)/4}
{exp}\left({-\frac{r^2}{8\theta}}\right)M_{W} \left( \frac{n+1}{4}
, \frac{n+3}{4}, \frac{r^2}{4\theta} \right)\right\}
\right]}{(n+3)\left(\frac{n+1}{2}\right)}
   \\+ \frac{16 \pi }{n r^{n-2}(4 \pi \theta)^{(n+1)/2}
}\frac{\left[ 2^{(3-n)/2} r^{n-1}\left(
\frac{1}{\theta}\right)^{(n-1)/2} \left(
\frac{r^2}{\theta}\right)^{-(n+1)/4}{exp}\left({-\frac{r^2}
{8\theta}}\right)M_{W} \left( \frac{n+5}{4},\frac{n+3}{4},
\frac{r^2}{4\theta} \right) \right]}{\frac{n+1}{2}}~~ +~~
\frac{C}{r^{n-2}}$
\begin{equation} \end{equation}

where $C$ is an integration constant and the Whittaker mass
$M_{W}$ can be defined as
\begin{eqnarray} M_{W} (\mu, \nu, z)
 = exp\left(-\frac{z}{2}\right)
z^{(1/2+\nu)}{hypergeom}\left( \left[\frac{1}{2}+\nu-\mu\right],
[1+2\nu], z \right).\end{eqnarray}

This solution is obviously very difficult to deal with. In the
following subsections we will therefore discuss various cases for
specific dimensions. However, $C$ being an integration constant,
mathematically, $b(r)$ is a solution for every values of $C$. The
flare-out condition, on the other hand, is a physical requirement
that is satisfied only for a certain range of values of $C$,
which, in turn, depend on the other parameters. Because of the
complexity of the shape function, we will use a graphical approach
to analyze the wormhole structure, including the location of the
throat, in terms of typical values of the parameters. In this
manner it is shown that wormhole solutions exist in four and five
dimensions only. The higher-dimensional cases require some
additional discussion and will be dealt with in the later phase.

\subsection{$n=2$}
When $n=2$, i.e., for a four-dimensional spacetime, the shape function
of the wormmhole takes the following form:

\begin{equation}
b(r) = \frac{M}{\sqrt{\pi \theta^3}}\left[ -2 \theta~ r ~{exp}
\left(-\frac{r^2}{4\theta}\right) + 2 \theta^{\frac{3}{2}}
\sqrt{\pi}~{erf}\left(\frac{r}{2\sqrt{\theta}}\right) +C\right].
\end{equation}

The other parameters are
\begin{eqnarray}
8 \pi p_r = -\frac{1}{r^3} \left[\frac{M}{\sqrt{\pi
\theta^3}}\left\{ -2 \theta~ r ~{exp}
\left(-\frac{r^2}{4\theta}\right) + 2 \theta^{\frac{3}{2}}
\sqrt{\pi} ~{erf}\left(\frac{r}{2\sqrt{\theta}}\right) +C\right\}
\right]
\end{eqnarray}

and

\begin{eqnarray}
8 \pi p_t = \frac{1}{2r^3} \left[\frac{M}{\sqrt{\pi
\theta^3}}\left\{ -2 \theta~ r ~{exp}
\left(-\frac{r^2}{4\theta}\right) + 2 \theta^{\frac{3}{2}}
\sqrt{\pi} ~{erf}\left(\frac{r}{2\sqrt{\theta}}\right) +C\right\}
\right]  \nonumber \\
-  \frac{8 \pi M}{(4 \pi
\theta)^{3/2}}~{exp} \left(-\frac{r^2}{4\theta}\right).
\end{eqnarray}

The next step is to verify that the shape function has all the
properties required for a wormhole structure.  To do so, we
need to assign some typical values to the parameters, an
example of which is shown in Fig. 1.  We can see from Fig. 2 that
$b(r)/r \rightarrow 0$ as $r\rightarrow \infty$, so that in
conjunction with the constant redshift function, the spacetime is
asymptotically flat.  The throat of the wormhole is located at
$r=r_0$, where $G(r)=b(r)-r$ cuts the $r$-axis, shown in Fig. 3.

Fig. 3 also indicates that for $r>r_0$, $G(r)<0$, i.e., $b(r)-r<0$,
which implies that $b(r)/r<1$ for $r>r_0$, an essential requirement
for a shape function. Moreover, $G(r)$ is a decreasing function for
$r\ge r_0$.  Since $G'(r)<0$, we have $b'(r_0)<1$, which is the
flare-out condition.  With this condition now satisfied, the shape
function has produced the desired wormhole structure.  For
completeness let us note that for the given parameters,
$r_0=0.1955$ to four decimal places with
$b'(0.1955)\approx 0.20$, obtained from Eq. (\ref{E:rho}).

The close connection between the flare-out condition and the
energy conditions calls for a check on the latter.  According to
Fig. 4, the null energy condition (NEC) is satisfied, but the weak
energy condition (WEC) and strong energy condition (SEC) are
violated.

 \begin{figure}
        \includegraphics[scale=.30]{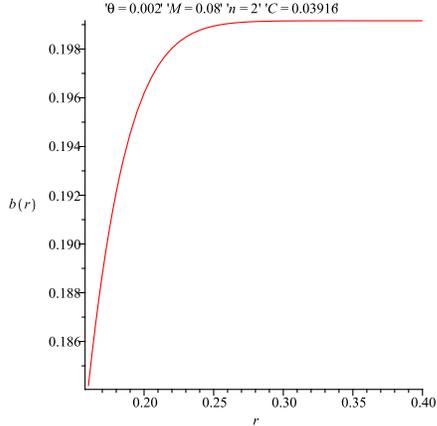}
        \caption{Diagram of the shape function of the wormhole in four dimension for the specific values of the parameters
        as $\theta =0.002$, $M=0.08$ and $C=0.03916$. }
   \label{fig:shape2}
\end{figure}
\begin{figure}
        \includegraphics[scale=.30]{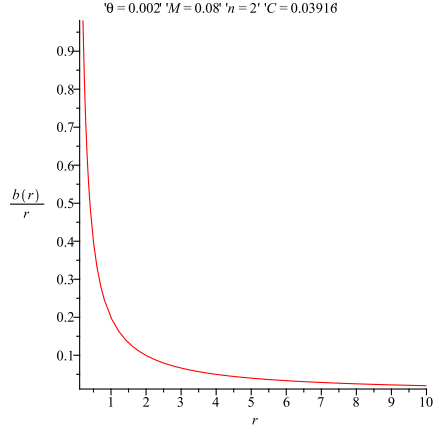}
        \caption{Asymptotic behavior  of the shape function of the wormhole given in Fig.1.}
   \label{fig:shape2}
\end{figure}
\begin{figure}
        \includegraphics[scale=.30]{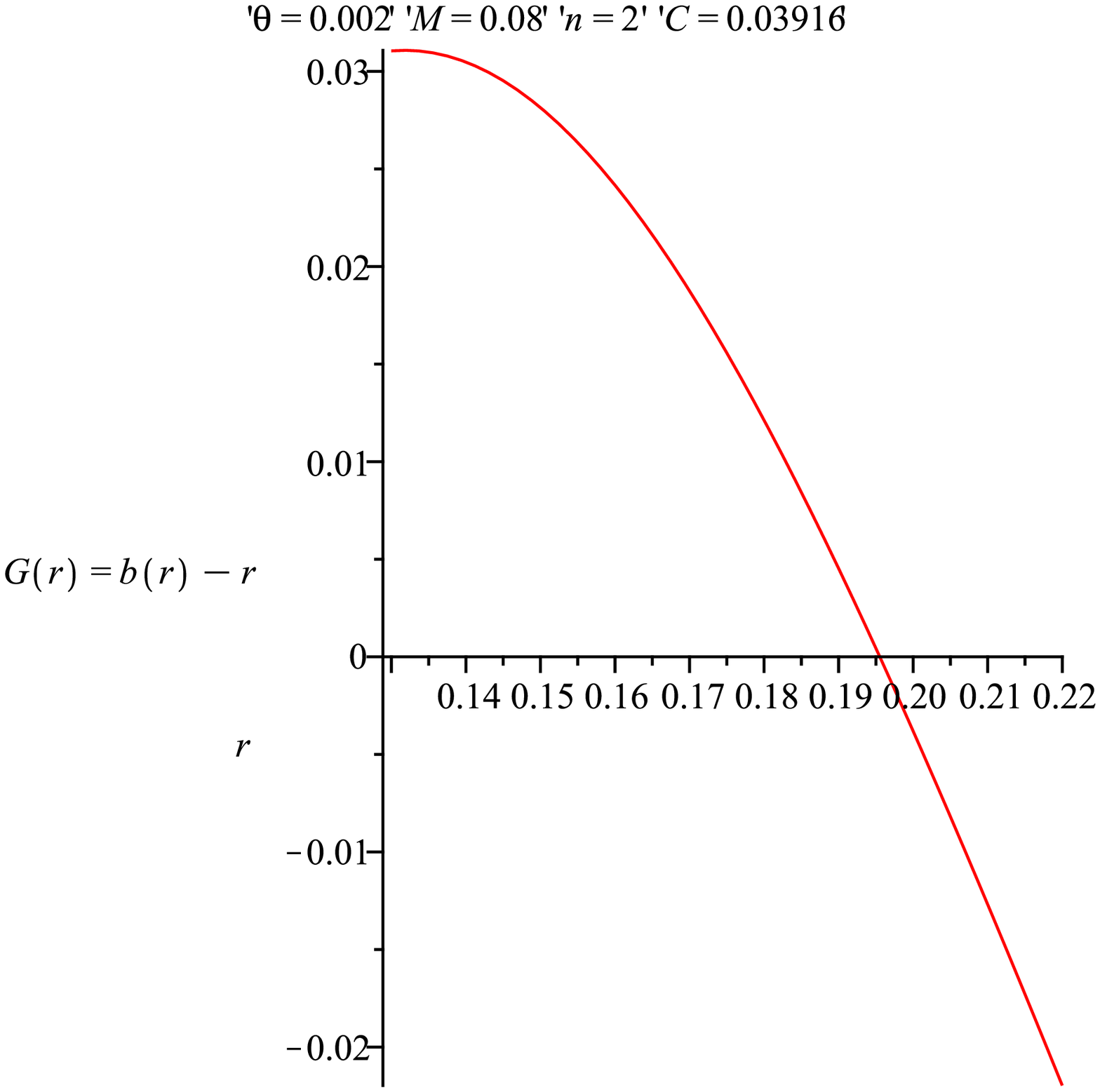}
        \caption{The throat of the wormhole given in Fig.1, occurs where $G(r)$ cuts the $r$-axis.}
   \label{fig:wh20}
\end{figure}
\begin{figure}
        \includegraphics[scale=.30]{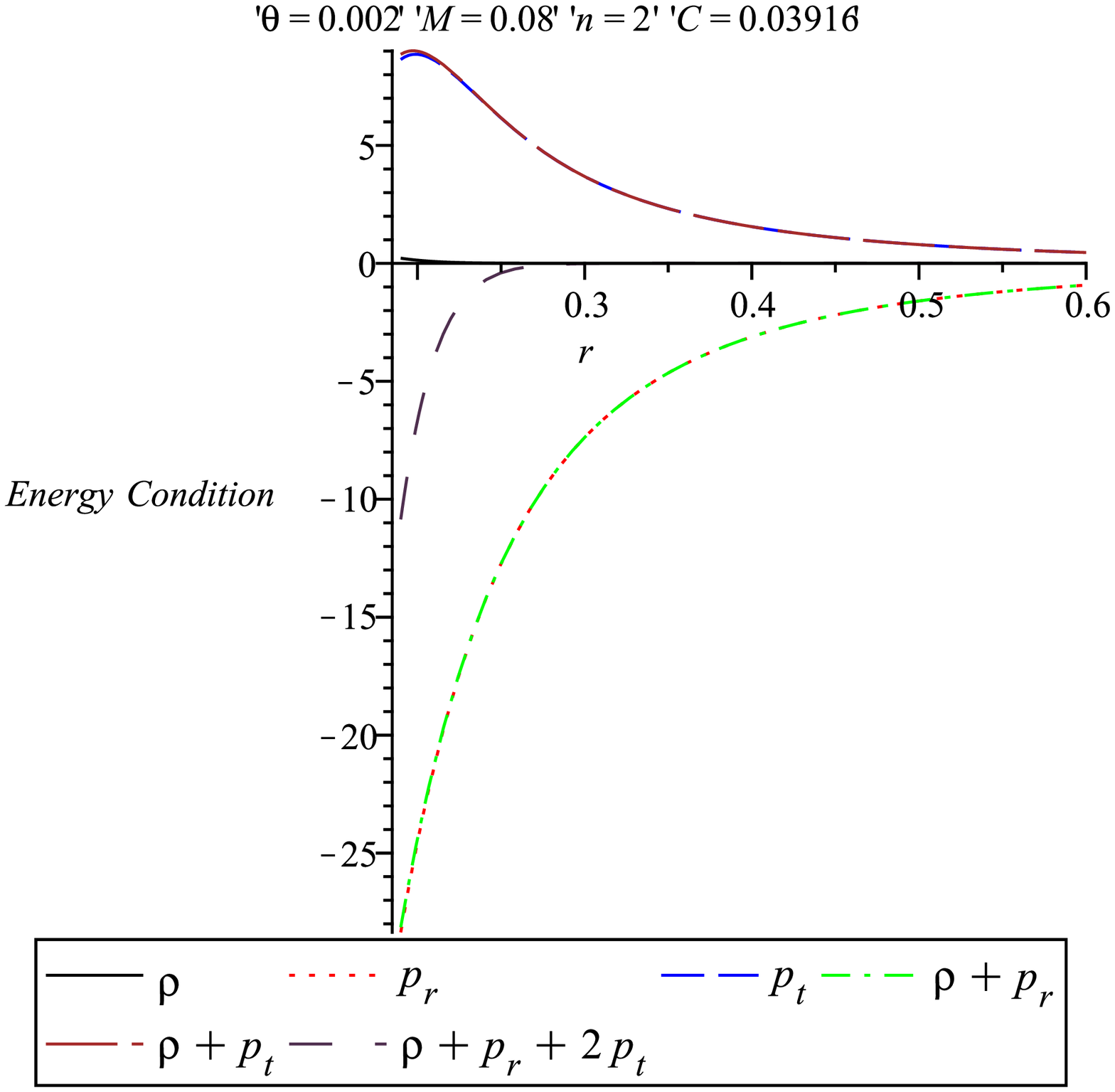}
        \caption{The variations of the left-hand sides of the expressions
for the energy conditions of the matters comprising the four
dimensional wormhole are plotted against $r$.}
   \label{fig:wh20}
\end{figure}

\subsection{$n=3$}
When $n=3$, i.e., for a five-dimensional spacetime, the shape function
of the wormmhole is given by

\begin{equation}
b(r) = \left[ \frac{-2\theta(4\theta +r^2)~M~ {exp}
\left(-\frac{r^2}{4\theta}\right)}{3 \pi r \theta^2}\right] +
\frac{C}{r}.
\end{equation}

The other parameters are

\begin{equation}
8 \pi p_r = -\frac{3}{r^3} \left\{\left[ \frac{-2\theta(4\theta
+r^2)~M~ exp \left(-\frac{r^2}{4\theta}\right)}{3 \pi r
\theta^2}\right] + \frac{C}{r} \right\}
\end{equation}
and

\begin{eqnarray}
8 \pi p_t = \frac{1}{r^3} \left\{\left[ \frac{-2\theta(4\theta
+r^2)M{exp} \left(-\frac{r^2}{4\theta}\right)}{3 \pi~ r~
\theta^2}\right] + \frac{C}{r} \right\}  \nonumber \\ -  \frac{16
\pi M}{3 (4 \pi \theta)^2}~{exp}
\left(-\frac{r^2}{4\theta}\right).
\end{eqnarray}

One can see from Fig. 5 that $b(r)=r$ at some point. More
precisely, for the particular parameters chosen, $b(0.0878)=
0.0878$ to four decimal places.  This is in agreement with Fig. 6,
which shows $G(r)$ intersecting the $r$-axis at $r=r_0$.  Also,
from Eq. (\ref{E:rho}), $b'(0.0878)\approx 0.34<1$.  The function
continues to increase to the right of $r_0=0.0878$; its slope and
is still positive at $r_1=0.0893$ but soon becomes negative. So we
have a valid solution around the throat.  Moreover, $r_1=0.0893$
is a convenient cut-off for the wormhole material and subsequent
junction to the corresponding external Schwarzschild spacetime
\[
  ds^2=-\left(1-\frac{2\mu}{r^2}\right)dt^2+\left(1-
    \frac{2\mu}{r^2}\right)^{-1}dr^2+r^2
    (d {{{\theta}_1} ^2} + {sin}^2 {\theta}_1 d
{{{\theta}_2} ^2}+ {sin}^2 {\theta}_1 {sin}^2 {\theta}_2 d
{{{\theta}_3} ^2}).
 \]
Here $\mu $ is related to the mass $`m'$ of the five dimensional
Schwarzschild black hole as $\mu = \frac{4 G m}{3 \pi}$.
Therefore, for the present case it can be evaluated as
$\mu=\frac{1}{2}r_1b(r_1)\approx 0.0039318$, while
$e^{\nu}=\left(1-\frac{2\mu}{r_1^2}\right)$, so that $\nu\approx
{ln}\left(1-\frac{0.0078636}{r_1^2}\right)\approx -4.27546$. This
yields the respective interior and exterior line elements:
\begin{equation}
  ds^2=-e^{-4.27546}dt^2+\frac{dr^2}{\left(1-\frac{b(r)}{r}\right)}
  +r^2( d {{{\theta}_1} ^2} + {sin}^2 {\theta}_1 d
{{{\theta}_2} ^2}+ {sin}^2 {\theta}_1 {sin}^2 {\theta}_2 d
{{{\theta}_3} ^2})
\end{equation}
for $r<r_1$ and
\begin{eqnarray}
  ds^2=-\left(1-\frac{0.0078636}{r^2}\right)dt^2
     +\frac{dr^2}{\left(1-\frac{0.0078636}{r^2}\right)}
 \nonumber \\ ~~~~~~~~~~~~~~~~~~~+r^2( d {{{\theta}_1} ^2} + {sin}^2 {\theta}_1 d
{{{\theta}_2} ^2}+ {sin}^2 {\theta}_1 {sin}^2 {\theta}_2 d
{{{\theta}_3} ^2})
\end{eqnarray}
for $r\ge r_1$.  (Due to the spherical symmetry, the remaining
components are already continuous \cite{LLS03}). Observe that only
the interior solution retains the zero tidal forces. One can note
that here the junction is a thin shell. Since the shell is
infinitely thin, the radial pressure is zero. According to the
Darmois-Israel junction condition \cite{IS1966}, the surface
density is also zero, however, the tangential pressure is nonzero
(see Ref. \cite{FR2006} for details).

\begin{figure}
        \includegraphics[scale=.30]{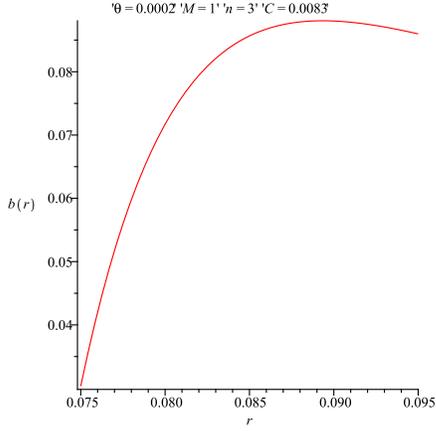}
        \caption{Diagram of the shape function of the wormhole in five dimension for the specific values of the parameters
        as $\theta =0.0002$, $M=1$ and $C=0.0083$. .}
   \label{fig:shape2}
\end{figure}
\begin{figure}
        \includegraphics[scale=.30]{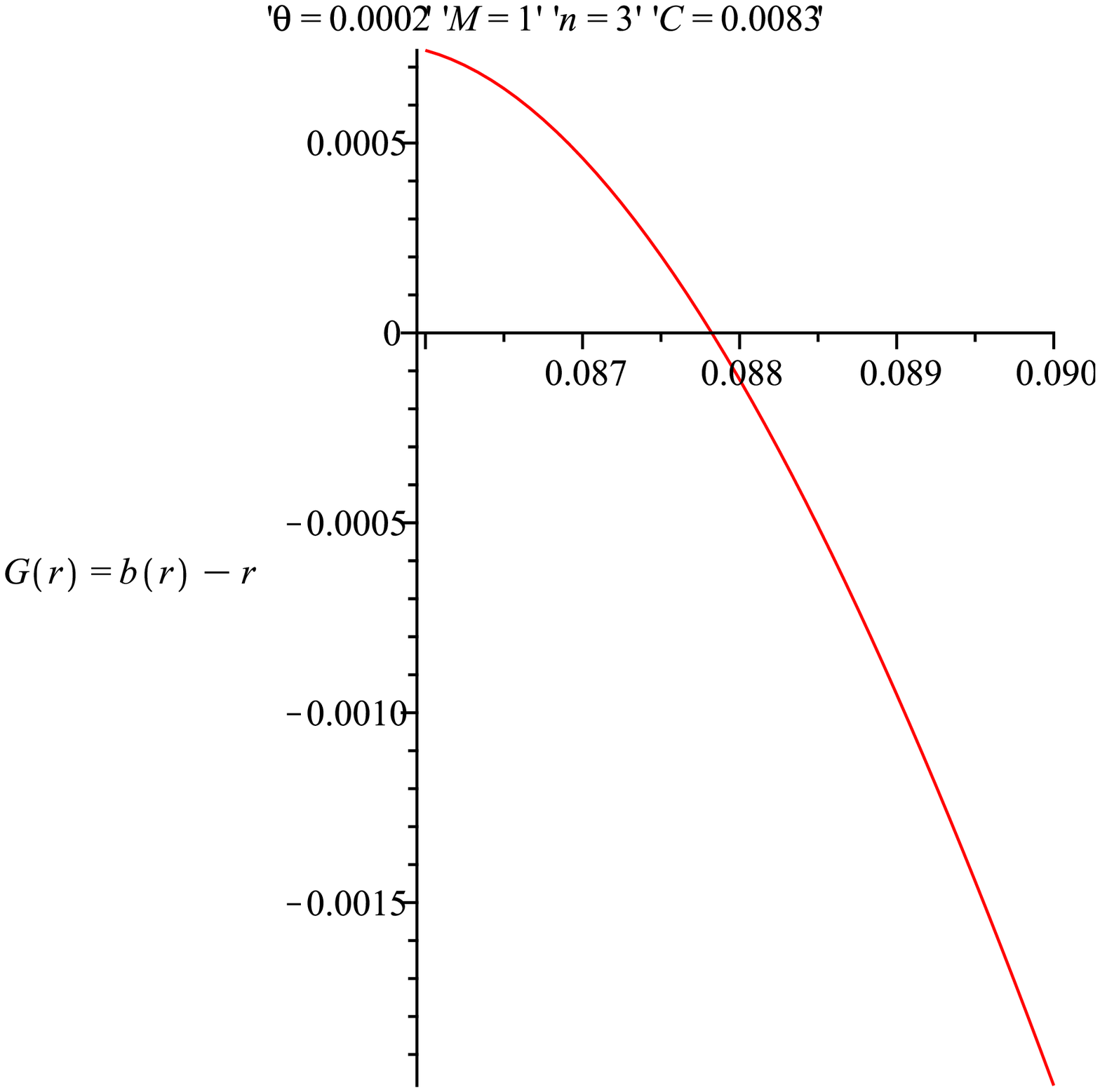}
        \caption{The throat of the wormhole given in Fig.5, occurs where $G(r)$ cuts the $r$-axis.}
   \label{fig:wh20}
\end{figure}
\begin{figure}
        \includegraphics[scale=.30]{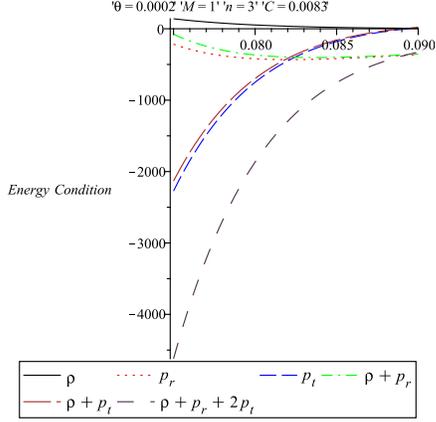}
        \caption{The variations of left-hand sides of the expressions
for the energy conditions of the matters comprising the five
dimensional wormhole are plotted against $r$.}
   \label{fig:wh20}
\end{figure}

For the five dimensional case, Fig. 7 indicates that only the NEC
is satisfied, while the WEC and SEC are violated.

\subsection{$n=4$ and $n=5$}
When $n=4$, i.e., for a six-dimensional spacetime, the shape function of
the wormmhole is

$ b(r)  = \frac{M}{8 \pi r^2 \sqrt{\pi \theta^5}} \left[ -2r^3
\theta~ {exp} \left(-\frac{r^2}{4\theta}\right) + 6 \theta
\left\{-2\theta ~r~ {exp} \left(-\frac{r^2}{4\theta}\right) + 2
\theta^{\frac{3}{2}} \sqrt{\pi}~ {erf}\left(\frac{r}{2
\sqrt{\theta}}\right)\right\}\right]$
\begin{equation}+
 \frac{C}{r^2}. \end{equation}

The other parameters are

$ 8 \pi p_r = -\frac{6}{r^3} \left[ \frac{M}{8 \pi r^2 \sqrt{\pi
\theta^5}} \left\{ -2r^3 \theta~{exp}
\left(-\frac{r^2}{4\theta}\right) + 6 \theta \left(-2\theta~ r
~e{xp} \left(-\frac{r^2}{4\theta}\right) + 2 \theta^{\frac{3}{2}}
\sqrt{\pi}~ e{rf}\left(\frac{r}{2
\sqrt{\theta}}\right)\right)\right\}\right] $
\begin{equation}- \frac{6 C}{r^5} ,\end{equation}

$ 8 \pi p_t =   \frac{3M}{16 \pi r^5 \sqrt{\pi \theta^5}} \left[
-2r^3 \theta~ e{xp} \left(-\frac{r^2}{4\theta}\right) + 6 \theta
\left\{-2\theta~ r ~{exp} \left(-\frac{r^2}{4\theta}\right) + 2
\theta^{\frac{3}{2}} \sqrt{\pi}~ e{rf}\left(\frac{r}{2
\sqrt{\theta}}\right)\right\}\right] $
\begin{equation}+ \frac{3C}{2r^5} - \frac{6
\pi M}{ (4 \pi \theta)^{\frac{5}{2}}}~{exp}
\left(-\frac{r^2}{4\theta}\right),
\end{equation}

Fig. 9 locates the presumptive throat of the wormhole and Fig. 8
shows the shape function $b(r)$. Observe that $b(r)$ is strictly
decreasing, a problem that also occurs in the case $n=5$ i.e. for
a seven dimensional spacetimes (see Fig. 10). For $n=5$, we get
\begin{eqnarray}
b(r)  = \frac{1}{r^3} \left[ C - \frac{(32 \theta^2 +8 \theta r^2
+r^4)M~{exp} \left(-\frac{r^2}{4\theta}\right)}{10 \pi^2
\theta^2}\right]
\end{eqnarray}

So it appears that there are no wormhole solutions for dimensions
above five as $b(r)$ contains the factor $\frac{1}{r^{n-2}}$ ($n
\geq 2$) which causes $b(r)$ to be a monotonic decreasing
function. The reason is that both ${exp}
\left(-\frac{r^2}{4\theta}\right)$ and $e{rf}\left(\frac{r}{2
\sqrt{\theta}}\right)$ level off rapidly in the outward radial
direction. As a result, $ b^\prime(r)$ is completely dominated by
$(\frac{d}{dr})(\frac{C}{r^{n-2}})$ for larger $n$ and proper
choice of $C$. So one would expect the slope of $b(r)$ to become
even steeper with increasing $n$, which is clearly born out in
Figs. 8 and 10. The steepness can only increase further as $n$
increases. For smaller $n$, this dominance does not necessarily
hold. Thus, according to Fig. 5, for $n=3$ and proper choice of
$C$, the term $\frac{C}{r}$ allows $b(r)$ to increase long enough
to yield an interior solution. For the case $n=2$, none of these
arguments apply, and the result is an asymptotically flat
spacetime.

\begin{figure}
        \includegraphics[scale=.30]{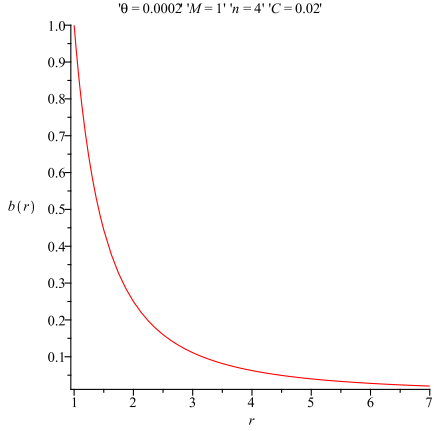}
        \caption{Diagram of the presumptive shape function of the wormhole  in six dimension for the specific values of the parameters
        as $\theta =0.0002$, $M=1$ and $C=0.02$. }
   \label{fig:shape2}
\end{figure}
\begin{figure}
        \includegraphics[scale=.30]{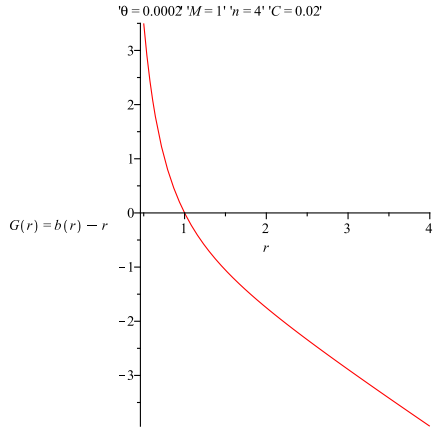}
        \caption{The throat of the wormhole given in Fig. 8 normally occurs where $G(r)$ cuts the $r$-axis.}
   \label{fig:wh20}
\end{figure}
\begin{figure}
        \includegraphics[scale=.30]{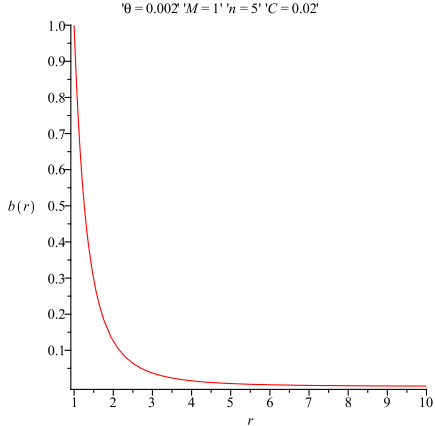}
        \caption{Diagram of the presumptive shape function of the wormhole in seven dimension for the specific values of the parameters
        as $\theta =0.0002$, $M=1$ and $C=0.02$. }
   \label{fig:shape2}
\end{figure}

\section{Conclusion}
Noncommutative geometry, an offshoot of string theory, replaces
point-like structures by smeared objects and has recently been
extended to higher dimensions. In this paper we obtain two
wormhole solutions within the framework of this extended
noncommutative geometry. In general, the primary ingredient for
sustaining traversable wormholes is the presence of exotic matter
that violates the null energy condition. However, we have proposed
a set of new wormhole solutions within the framework of
noncommutative geometry background, where the matter satisfies the
null energy condition (NEC) but violates the weak energy condition
(WEC) and strong energy condition (SEC). It is shown that wormhole
solutions exists in the usual four, as well as in five dimensions,
but they do not exist in six or seven dimensions, suggesting that
such solutions do not exist beyond five dimensions. This outcome
points out the danger of assuming only one extra spatial
dimension, as is often done. In the analysis of this work, we
considered a constant redshift function, the so-called zero-tidal
force solution  to make the wormhole traversable by humanoid
travelers. This assumption simplified the mathematical
calculations, but it also provided sufficiently exciting exact
solutions. It is worthwhile to mention here that the wormhole in
four dimensional spacetime is asymptotically flat whereas in five
dimension it is asymptotically non-flat and for higher than five
dimensions no wormhole exists. Thus up to four dimension, one can
get a regular wormhole and for five dimensional spacetime, one
gets wormhole geometry only in a very restricted region. The
possibility of getting a wormhole geometry will cease beyond five
dimensions.

\subsection*{Acknowledgments}
FR and SR are thankful to the Inter University Centre for
Astronomy and Astrophysics (IUCAA), India for providing research
facilities. FR is also grateful to UGC, India for financial
support under its Research Award Scheme. We would like to thank
the referee for his useful comments to improve substantially the
present work.

\end{document}